\documentclass[
aps,
amsmath,
amssymb,
floatfix,
twocolumn,
final,
prb,
aapm,
showpacs
]{revtex4-2}

\usepackage{dcolumn}
\usepackage{natbib}
\usepackage{url}

\usepackage{textcase}
\usepackage{bm}

\usepackage{float}
\usepackage{amsmath}
\usepackage{epsfig}
\usepackage{graphicx}
\usepackage[ansinew]{inputenc}
\usepackage{array}
\usepackage{hyperref}
\usepackage{tabularx}
\usepackage{color,soul}
\newcolumntype{C}[1]{>{\centering\let\newline\\\arraybackslash\hspace{0pt}}m{#1}}

\begin{document}

\preprint{APS/123-QED}
\title{Slow crystalline electric field fluctuations in the Kondo lattice SmB$_{6}$}

\author{M. Carlone$^{1}$, J. C. Souza$^{2,3}$, J. Sichelschmidt$^{3}$, P. F. S. Rosa$^{4}$, R. R. Urbano$^{2}$, P. G. Pagliuso$^{2}$, Z. Fisk$^{4}$, P. A. Venegas$^{5}$, P. Schlottmann$^{6}$ and C. Rettori$^{2}$}

\affiliation{$^{1}$ POSMAT-Programa de P\'os-Gradua\c{c}\~ao em Ci\^encia e Tecnologia de Materiais, Faculdade de Ci\^encias, Universidade Estadual Paulista-UNESP, Bauru, SP, CP 473, 17033-360, Brazil\\
$^{2}$Instituto de F\'isica \lq\lq Gleb Wataghin\rq\rq,Unicamp, 13083-859, Campinas, SP, Brazil\\
$^{3}$Max Planck Institute for Chemical Physics of Solids, D-01187 Dresden, Germany\\
$^{4}$Los Alamos National Laboratory, Los Alamos, New Mexico 87545, USA\\
$^{5}$Departamento de F\'isica, Universidade Estadual Paulista-Unesp, Caixa Postal 473, 17033-360 Bauru, SP, Brazil\\
$^{6}$Department of Physics, Florida State University, Tallahassee, FL 32306, USA}

\date{\today}

\begin{abstract}

This work reports on the temperature dependence of the electron spin resonance (ESR) of Gd$^{3+}$-doped SmB$_{6}$ single crystals at X- and Q-band microwave frequencies in different crystallographic directions. We found an anomalous inhomogeneous broadening of the Gd$^{3+}$ ESR linewidth ($\Delta H$) within 5.3 K $\leq T \leq$ 12.0 K which is attributed to slow crystalline electric field (CEF) fluctuations, slower than the timescale of the ESR microwave frequencies used ($\sim$10 GHz). This linewidth inhomogeneity may be associated to the coupling of the Gd$^{3+}$ $S$-states to the breathing mode of the SmB$_{6}$ cage, and can be simulated by a random distribution of the 4$^{th}$ CEF  parameter, $b_4$, that strikingly takes negative and positive values. The temperature at which this inhomogeneity sets in, is related to the onset of a continuous insulator-to-metal phase transition. In addition, based on the interconfigurational fluctuation relaxation model, the observed exponential $T$-dependence of $\Delta H$ above $T\simeq$ 10 K gives rise to an excitation energy notably close to the hybridization gap of SmB$_{6}$ ($\Delta\simeq$ 60 K). This charge fluctuation scenario provides important ingredients to the physical properties of SmB$_{6}$. We finally discuss the interplay between charge and valence fluctuations under the view of slow CEF fluctuations in SmB$_{6}$ by coupling the Gd$^{3+}$ ions to the breathing phonon mode via a dynamic Jahn-Teller-like mechanism.

\end{abstract}

\maketitle
\section{\label{sec:intro}Introduction}

The interplay of topology, strongly correlated electrons and/or magnetism may lead to new quantum states of matter, such as magnetic topological insulators \cite{mong2010antiferromagnetic}, Weyl-Kondo semimetals \cite{chang2018parity,Lai2018} and topological Mott insulators \cite{pesin2010mott}. The first topological phase of matter predicted in a strongly correlated $f$-electrons system was the topological Kondo insulator (TKI) that resulted in several experimental and theoretical works \cite{dzero2016topological}. In particular, SmB$_{6}$, a mixed valence compound with a hybridization gap $\Delta\simeq$ 60 K \cite{cooley1995sm,eo2018robustness}, is a heavily-studied material with disputed experimental results that either support or weaken the TKI scenario \cite{kim2014topological,kim2013surface,hlawenka2018samarium,Steven2013,Herrmann2020,li_emergent_2020}. In particular, conflicting results are observed when comparing experimental data from samples grown by different routes \cite{tan2015unconventional,li2014two,thomas2018quantum}. Nonetheless, it is essential to fully understand the bulk properties of SmB$_{6}$ in order to have a complete description of the possible gapless spin polarized surface states.

Two intriguing bulk features are the linear low-$T$ specific heat behaviour (Sommerfeld coefficient $\gamma$) \cite{wakeham2016low} and \lq\lq excitonic\rq\rq  states first observed with inelastic neutron scattering and Raman, and used to explain nuclear magnetic resonance (NMR) \cite{caldwell2007high,schlottmann2014nmr} and then supported by muon spin relaxation ($\mu$SR) experiments \cite{biswas2014low,biswas2017suppression}. The lowest reported gamma value is observed in double enriched samples, $\gamma$ = 2 mJ/mol K$^{2}$, but the $\gamma$ value in different growths may vary from $\gamma$ = 5 mJ/mol K$^{2}$ to about 25 mJ/mol K$^{2}$ \cite{valentine2018effect,Orend2017}. Moreover, recent specific heat measurements on SmB$_{6}$ single crystals and powdered samples showed that such finite $\gamma$ is not related to surface effects but rather of bulk origin \cite{wakeham2016low}. Such observations were interpreted in terms of a chargeless \lq\lq neutral\rq\rq~ Fermi surface in the bulk, \cite{schlottmann2014nmr,tan2015unconventional,erten2017skyrme} which may result in a \lq\lq failed superconductor\rq\rq \cite{erten2017skyrme}. Those particles couple to a magnetic field, but not to an electric field. Nevertheless, natural disorder in real materials could also contribute to the thermodynamic properties of the system \cite{sen2018fragility,Shen2018}.

The neutral particles scenario finds some support in NMR and $\mu$SR experiments. By investigating the $^{11}$B $\pm$3/2 $\leftrightarrow$ $\pm$1/2 transitions, previous NMR studies reported a deviation from an exponential-like activation behavior of the spin-lattice relaxation rate, 1/$T_{1}$, at low temperatures (6 K $\leq T \leq$ 15 K), \cite{caldwell2007high,takigawa1981nmr} which evolves and disappears as function of applied magnetic field. This deviation was interpreted as a result of excitonic in-gap states, namely magnetic in-gap bound states where antiferromagnetic correlations provide a plausible mechanism for the formation of the magnetic excitons \cite{schlottmann2014nmr,riseborough_9_2000}. In a similar $T$-range, low field ($H\leq$ 500 Oe) $\mu$SR experiments have shown a bulk dynamic magnetic field of magnitude $\Delta B$ = 18 Oe and slow timescale of $t$ = 60 ns \cite{biswas2014low,biswas2017suppression}. These authors have also shown that the bulk dynamic magnetic fields are suppressed nearby the surface for SmB$_{6}$ single crystals grown by the floating-zone technique. Therefore, Gd$^{3+}$ electron spin resonance (ESR) experiments, as a local probe in Gd-doped SmB$_{6}$ single crystals, may contribute to elucidate some of these intriguing issues. Gd$^{3+}$ couples via ferromagnetic Heisenberg exchange to the conduction electrons, $ce$. The antiferromagnetic Schrieffer-Wolff exchange is much smaller, so that the Gd$^{3+}$ ions do not display Kondo effect. Only the valence fluctuations of SmB$_{6}$ show a Kondo-like behavior.

Hence, this work aims to shed light into this puzzle by studying high quality Sm$_{1-x}$Gd$_{x}$B$_{6}$ single crystalline samples grown by the flux method, by means of the ESR microscopic technique at different frequencies (X band: 9.4 GHz and Q band: 34.1 GHz) in the temperature range of 2.5 K $\leq T\leq$ 20 K. A highly diluted Gd$^{3+}$-doping regime was chosen ($x$ = 0.0004) to preclude undesired Gd$^{3+}$-Gd$^{3+}$ magnetic interactions. 

\section{\label{sec:experiment}Experimental Details}

Single crystalline samples of Sm$_{1-x}$Gd$_{x}$B$_{6}$ were synthesized by Al-flux grown technique as described elsewhere \cite{eo2018robustness,RosaFisk2018}. The crystals used in our experiments had typical dimensions of $\sim 700\mu$m $\times$ 300$\mu$m $\times$ 120$\mu$m and estimated skin depth $\delta \simeq 100 \mu$m at $T\leq$ 4 K ($\delta =\sqrt{\rho /\pi \nu \mu }$). Magnetic susceptibility measurements were carried out in a SQUID vibrating-sample magnetometer. ESR measurements in as-grown facets of cubic single crystals were performed in X- (9.4 GHz) and Q-bands (34.1 GHz) spectrometers equipped with a goniometer and a He-gas flow cryostat able to vary the temperature within the range of 2.5 K $\leq T\leq$ 20 K. The studied single crystals were not polished but etched before the ESR measurements in a 3:1 mixture of hydrochloric and nitric acids (aqua regia) to remove residual impurities on their surfaces due to the Al flux. The studied sample masses ranged from 0.3 mg to 4.0 mg. Along this report, the concentration $x$ refers to the nominal concentration value \cite{souza2020metallic}.

\section{\label{sec:TheoreticalModel}Theoretical Model}

The absorption response with exchange narrowing effects due to the exchange interaction between the Gd$^{3+}$ local moment and the $ce$, is obtained by calculating the transverse dynamic susceptibility $\chi^{+}(\omega)$. Assuming no bottleneck relaxation effects,  $\chi^{+}(\omega)$ can be approximated by \cite{Plefka73,Urban75,venegas2016collapse,duque2007exchange}:  
\begin{equation}
	\chi^{+}(\omega) \approx 1 -\omega_0  \sum_{M',M}^{} \left[ P_M(\Omega^{-1})_{M,M'}   \right],
	\label{Eq1} 
\end{equation}
where $P_M$ are the transition probabilities between the $M \leftrightarrow M+1$ states, and can be written as: 
\begin{equation}
	P_M = C_M \; exp^{M \hbar\omega_0/kT} /\; \sum_{M'} C_{M'} \; exp^{M' \hbar\omega_0/kT}.    
	\label{Eq2} 
\end{equation}

The coefficients $C_M$ are defined as $C_M = S(S+1)-M(M+1)$, with $M$ and $M'$ the quantum numbers associated with the Gd$^{3+}$ ($S$ = 7/2) Zeeman split states. $\Omega_{M,M'}$ is the transition matrix, which can be written as:
\begin{equation}
	\begin{split}
		\Omega_{M,M'} = [H_0 -H -H_M]\delta_{M,M'} - i\delta_{M,M'} \Delta H_{res} \\ - \; i(1/2){\Delta H}{C_{M'}}[ 2\delta_{M,M'} - \delta_{M,M'+1} - \; \delta_{M,M'-1} ].  
	\end{split}
	\label{Eq3} 
\end{equation}

 The real part of the diagonal elements contains the resonance field, and the remaining imaginary term corresponds to the ESR relaxation of Gd$^{3+}$ ions.  In this equation, $H_{0}=\hbar\omega_0/g\mu_B$ with $\omega_0$ being the microwave frequency, and $\Delta H_{res}$ the residual linewidth. $\Delta H$ is associated to $T$-dependent relaxation processes, where the diagonal terms are responsible for the ESR linewidth, and the off-diagonal elements contain the fluctuation rates of the local moments between two consecutive resonance fields, and are responsible for the narrowing effects of the fine structure.
 
 The Gd$^{3+}$ ESR spectrum in an insulating host presents a fine structure of seven Lorentzian resonances due to the crystal electric field (CEF) on Gd$^{3+}$ ions located at cubic sites. The spectrum is then calculated using the well-known CEF Hamiltonian for cubic symmetry in the presence of a magnetic field \cite{barnes1981theory,abragam2012electron}:
\begin{equation}
	\mathcal{H} = g\mu_B \textbf{H} . \textbf{S} + (1/60) b_4  \left(O{^0_4} +  5O{^4_4} \right),    
	\label{Eq4} 
\end{equation}
where the first term is the Zeeman interaction with $\mu_B$ the Bohr magneton, $b_{4}$ the fourth order CEF parameter, and $O^{m}_{4}$ the fourth order Stevens' operators. The contribution of the sixth order CEF term is negligible and has not been considered for simplicity.

In a cubic environment, the fine-structure resonance fields, including their angular dependence,  are:
\begin{equation}
	H_M = (1/60) b_4 p(\theta)\left<M|O{^0_4} + 5O{^4_4}|M \right> ,   
	\label{Eq5} 
\end{equation}
with $p(\theta) = 1 -5[sin^2 \theta - (3/4)sin^4 \theta ]$ being the angular dependence for cubic symmetry, and $\theta$ the angle between the applied magnetic field and the crystallographic axis. In this work, the applied magnetic field $H$ is always rotated within the (110) plane. 

However, in a metallic host, the absorption is given by the real part of the impedance, that is ${P \propto Re(\chi^{+} (\omega)) - Im(\chi^{+}(\omega)) }$ \cite{CKittel87,Urban75}. The transverse dynamic susceptibility $\chi^{+}$ is obtained from Eq. (\ref{Eq1}), that includes all the narrowing effects of the fine structure. The  elements of the transition matrix are calculated using Eqs. (\ref{Eq4}) and (\ref{Eq5}).  The linewidth and $g$-value are extracted from the simulated spectra using Feher's formula ${P \propto [\xi \chi'  + (\xi -1) \chi'']}$ \cite{Bloembergen1952,Dyson1955,Feher1955}, where  0 $\leq \xi \leq$ 1 . For $\xi=1$ we have a pure Lorentzian line, and for $\xi<1$ we have the well known Dysonian asymmetric ESR resonance lineshape.

 Using Eqs. (\ref{Eq1}) to (\ref{Eq5}) it is possible to fully describe the fine structure of Gd$^{3+}$, including the exchange narrowing effects. However, in order to describe the inhomogeneous broadening of each resonance of the fine structure, slow fluctuations of the CEF  (slower than $\sim$10 GHz, the frequency of our ESR experiments) will be considered through a $T$-dependent Gaussian distribution of the CEF $b_4$ parameter (GDCEF) with standard deviation $\sigma_{b_4}$. The origin for these slow CEF fluctuations will be associated to the coupling of the Gd$^{3+}$ $S$-states to the breathing mode of the SmB$_{6}$ cage (see below) \cite{lesseux_anharmonic_2017, Jolanta2021}.
 
\section{\label{sec:resultsanddiscussion}Results and Discussion}

\begin{figure}[!t]
	\includegraphics[width=0.9\columnwidth]{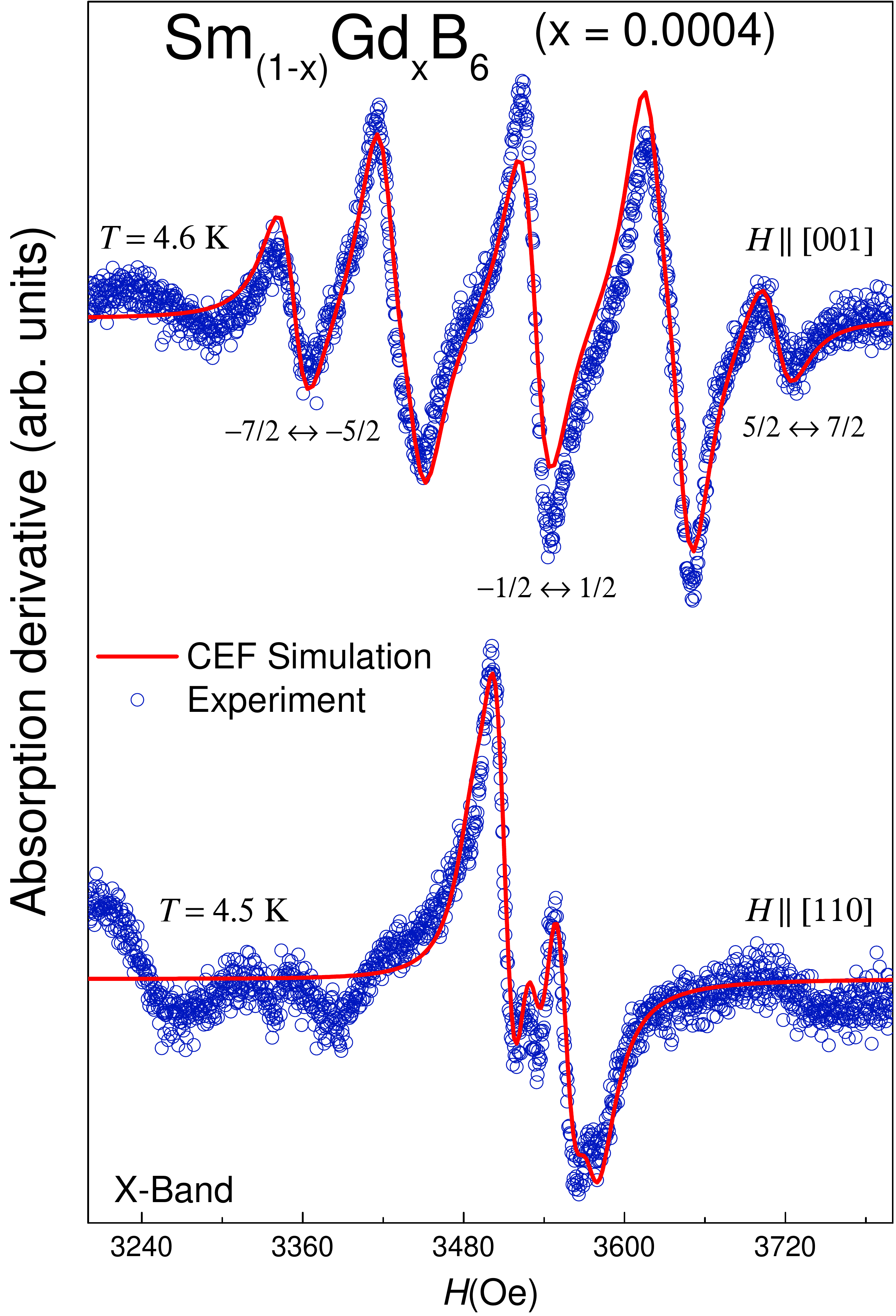}
	\caption{X-band Gd$^{3+}$ ESR spectra for Sm$_{0.9996}$Gd$_{0.0004}$B$_{6}$ with $H\parallel [001]$ and $H\parallel [110]$ at nearly 4.5 K \cite{souza2020metallic}. The solid red lines are simulations using Eqs. (\ref{Eq4}) and (\ref{Eq5}) considering Gd$^{3+}$ ions in a cubic insulating matrix without the influence of the GDCEF.}
	\label{Fig01}
\end{figure}

Figure \ref{Fig01} shows the X-band (9.4 GHz) Gd$^{3+}$ experimental ESR spectra for Sm$_{0.9996}$Gd$_{0.0004}$B$_{6}$ with the external magnetic field $\textbf{$H_0$}$ applied along the [001] and [110] directions at 4.6 K and 4.5 K, respectively. The red solid lines in Figure \ref{Fig01} are the simulated Gd$^{3+}$ fine structure of the spectra for both cases using seven Lorentzian lines. Because the skin depth is of the order of the crystal dimensions at these temperatures, it is natural that the lineshape of each individual resonances resembles an insulating host. Even at 34.1 GHz (Q-band) it was not possible to capture the skin depth conductivity effects in this sample \cite{PhysRevB.94.165154,LAURITA201878,souza2020metallic}. Note that at lower $x$ = 0.0002 \cite{Souza2021-2} and higher $x$ = 0.02 \cite{souza2020metallic} also diffusive-like and Dysonian lineshapes were reported, respectively. The obtained CEF parameters are roughly the same in both cases: $b_{4}$ = (- 9.9 $\pm$ 1.2) Oe; $g$ = 1.820 for $H\parallel$ [001] and $b_{4}$ = (- 9.0 $\pm$ 1.2) Oe; $g$ = 1.821 for $H\parallel$ [110]. Besides, quite narrow residual linewidths were considered in these simulations, $i.e.$, $\Delta H_{res}^{[001]}$ = (19.5 $\pm$ 1.0) Oe and $\Delta H_{res}^{[110]}$ = (17.4 $\pm$ 0.9) Oe. 

 In order to study the $T$-evolution of the ESR fine structure in a mixed valence insulator, one should also consider the $T$-dependent relaxation process due to the valence fluctuation (VF) between the $4f^{n}$ and $4f^{n+1}$ configurations of Sm ions. This causes a fast fluctuating field at the Gd$^{3+}$ site \cite{gambke1978epr,Rosa2020,venegas1992epr,venegas2016collapse} which homogeneously broadens the linewidth of the individual resonances leading to the narrowing of the Gd$^{3+}$ CEF fine structure. 

\begin{figure}[!b]
	\includegraphics[width=0.95\columnwidth]{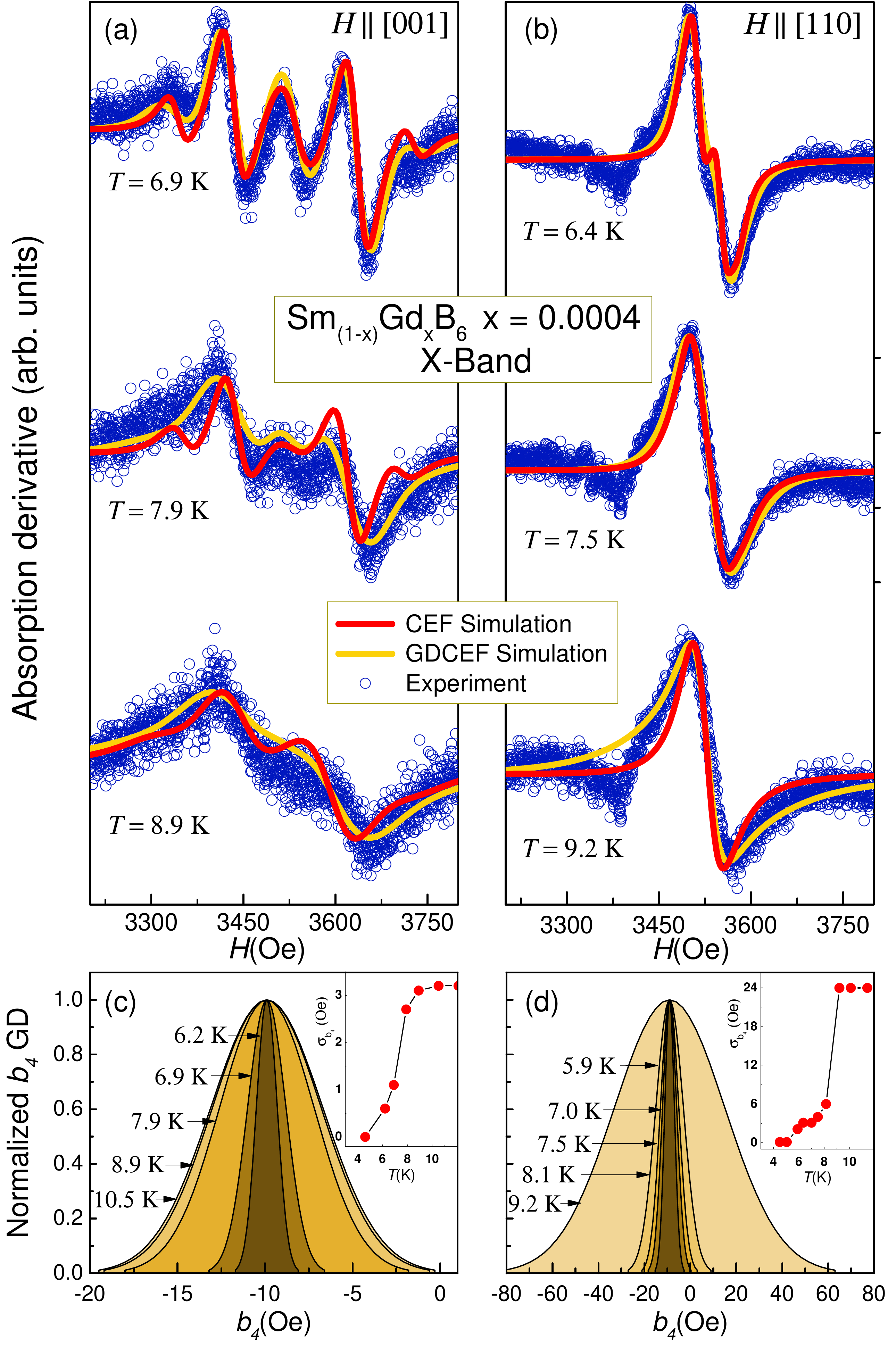}
	\caption{$T$-dependence of X-band Gd$^{3+}$ ESR spectra for (a) $H\parallel [001]$ and (b) $H\parallel [110]$. The yellow and red solid lines are the best simulations of the Gd$^{3+}$ ESR spectra with and without the influence GDCEF, respectively. (c) and (d) present the GDCEF for both cases and the insets show the $T$-dependence of their standard deviations $\sigma_{b_4}$.}
	\label{Fig02}
\end{figure}

\begin{figure}[!b]
	\includegraphics[width=0.95\columnwidth]{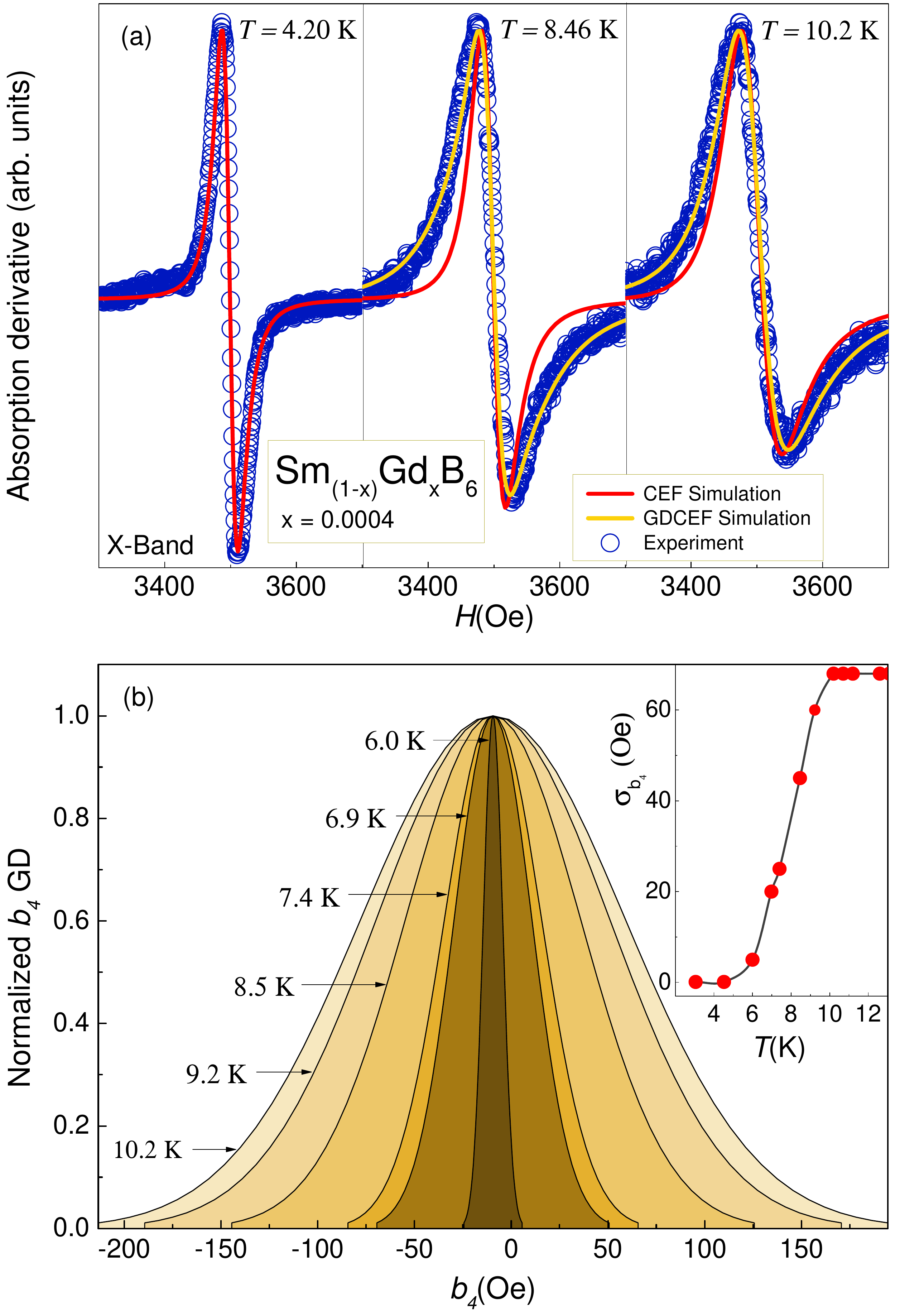}
	\caption{(a) X-Band $T$-dependence of the Gd$^{3+}$ ESR lineshapes at $H\parallel 30^\circ$: nearly the angle of collapsed fine structure. The yellow and red solid lines are simulations with and without GDCEF, respectively. All three spectra are on the same scale; (b) Gaussian distribution of $b_{4}$ (GDCEF) used to simulate the nearly collapsed ESR spectra at different temperatures. The inset shows the $T$-dependence of the GDCEF standard deviation $\sigma_{b_4}$.}
	\label{Fig03}
\end{figure}

We have calculated the individual linewidth $\Delta H(T)$, at any orientation, using a $T$-independent residual linewidth and homogeneous $T$-dependent contributions due to Korringa and Sm VF relaxation mechanism processes, as follows \cite{venegas1992epr}:
\begin{equation}
\Delta H = \Delta H_{res} + bT + Ae^{-E_{ex}/T},
\label{Eq06}
\end{equation}
where the first term is the residual linewidth and $b$ the usual Korringa relaxation rate \cite{Korringa1950} associated to a possible relaxation process due to the exchange interaction, $J_{fs}$, between the Gd$^{3+}$ localized magnetic moment and the $ce$. The last term is the exponential contribution to the Gd$^{3+}$ relaxation due to the Sm$^{2.6+}$ VF \cite{venegas1992epr}, where $A$ is a constant and $E_{ex}$ is the interconfigurational excitation energy, $i.e$., the energy necessary to exchange an electron between the $ce$-band and the Sm $4f$ states \cite{venegas1992epr,venegas2016collapse}, actually, the energy to excite an electron from the hybridized valence band into the hybridized conduction band (indirect gap).  Figures \ref{Fig02}(a) and \ref{Fig02}(b) show the $T$-evolution of the X-band Gd$^{3+}$ ESR spectra shown in Figure \ref{Fig01}, where we have used Eq. (\ref{Eq06}) to describe the $T$-evolution of the linewidth for each component of the fine structure. The parameters obtained for the spectra in Figure \ref{Fig01} were further confirmed by the ESR spectra taken at $H$ nearly $30^\circ$ from the [001] direction in the (110) plane where the cubic fine structure is about to collapse, as shown in Figure \ref{Fig03}(a) at 4.2 K. These simulations were able to capture most of the expected details of the cubic CEF fine structure of the ESR spectra for quite isolated Gd$^{3+}$ ions in SmB$_{6}$, confirming the high dilution of our single crystals.

\begin{figure}[!b]
	\includegraphics[width=0.95\columnwidth]{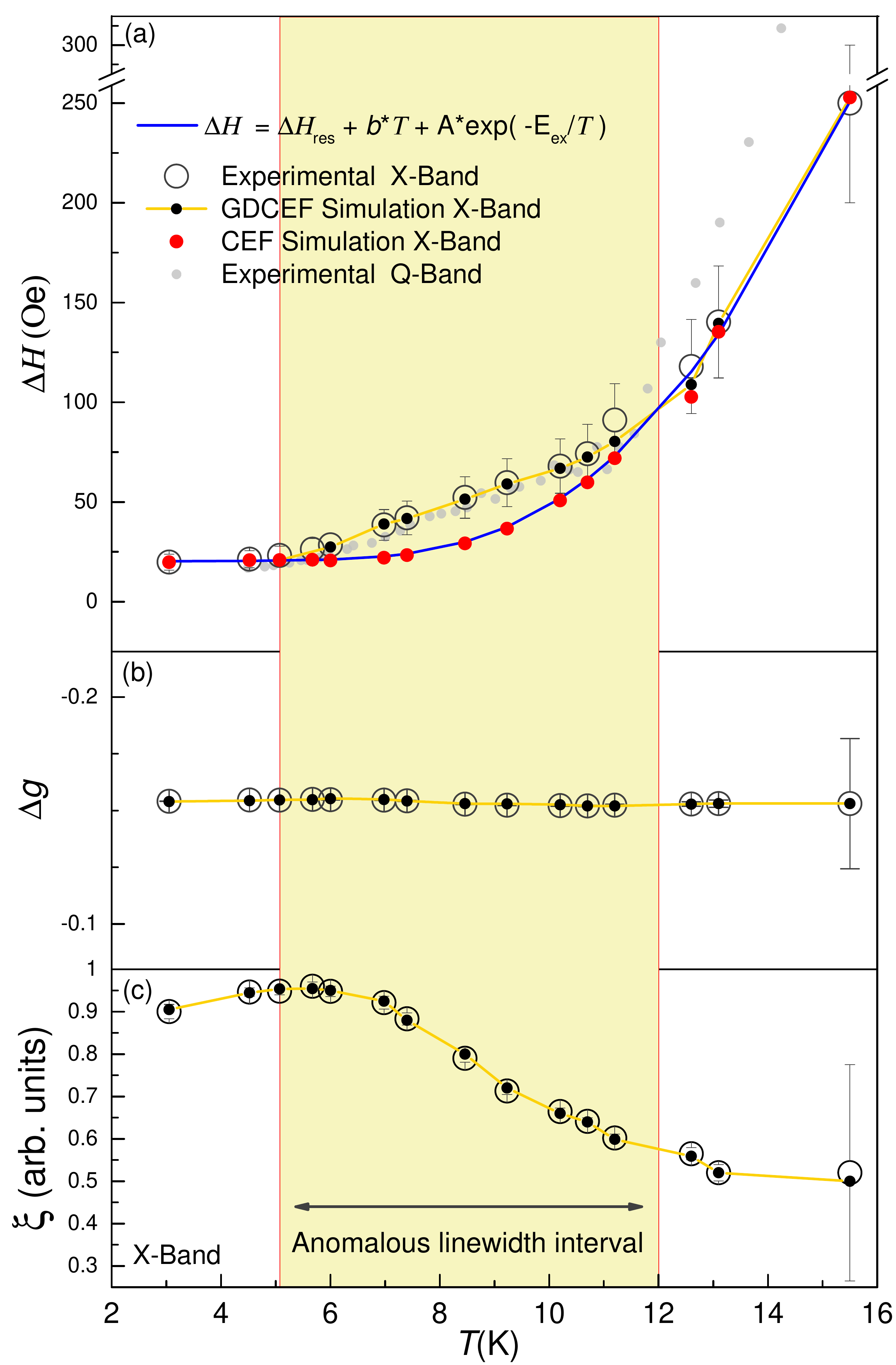}
	\caption{(a) $T$-dependence of the Gd$^{3+}$ ESR linewidth for X (empty black circles) and Q (solid gray dots) bands, at $H\parallel 30^\circ$: the angle of nearly collapsed fine structure. The theoretical simulations with GDCEF (solid black dots and yellow solid line), without GDCEF (solid red dots) and by Equation (\ref{Eq06}) (solid blue line). The used parameters were: $\Delta{H}_{res}$ = 20 Oe, $E_{ex}$ = 60 K, and above the insulator-to-metal transition, a Korringa rate of $b$ = 0.1 Oe/K; (b) the $g$-shift, $\Delta g$, and $\xi$ as a function of temperature, respectively. The metallic character of the resonance lines is evident in (c). The yellowish band, on the background, highlights the anomalous linewidth interval (5.3 K $\leq T \leq$ 12.0 K). }
	\label{Fig04}
\end{figure}

Therefore, SmB$_{6}$ behave as an insulator at low temperatures ($T\leq$ 6 K) and the Gd$^{3+}$ ESR spectra are well described by Eq. (\ref{Eq4}), with narrow residual linewidths, $\Delta H_{res}$. Yet at the insulator-to-metal crossover temperature and above, due to the presence of $ce$ and Sm VF, one must also take into account contributions from the Korringa and Sm VF relaxation processes to the ESR linewidth. Nevertheless, at intermediate temperatures (5.3 K $\leq T \leq 12$ K) highlighted in Figure \ref{Fig04}, these contributions were not able to simulate the experimental ESR spectra satisfactorily, as shown by red solid lines in Figures \ref{Fig02}(a) and \ref{Fig02}(b).

In order to account the discrepancy between experiment and calculation, an inhomogeneous linewidth broadening through the GDCEF, with a $T$-dependent standard deviation, $\sigma_{b_4}$, was considered in our simulations. The yellow solid lines in Figures \ref{Fig02}(a) and \ref{Fig02}(b) are the best simulations of the Gd$^{3+}$ ESR spectra caused by this GDCEF. The improvement of the agreement between simulation and data is noticeable. Figures \ref{Fig02}(c) and \ref{Fig02}(d) show the $T$-dependence of the GDCEF used to simulate the spectra of Figures \ref{Fig02}(a) and \ref{Fig02}(b), respectively. Notice that the $b_{4}$ CEF parameter distribution, which reproduce all the features of the experimental spectra, depends on the field direction.  

The presence of an inhomogeneous distribution of $b_{4}$ can be further confirmed by a thorough analysis of the Gd$^{3+}$ ESR spectra for the field direction where the CEF fine structure is fully collapsed into one single resonance line. For cubic symmetry this situation occurs at $\theta=29.67^{\circ}$ from the [001] direction in the (110) plane, i.e. $p(\theta)$=0 in Eq. (\ref{Eq5}). At this angle there is no contribution to the position of the spectra from CEF effects. Hence, the seven Gd$^{3+}$ ESR resonances overlap at the resonance field of the ($+1/2 \leftrightarrow -1/2$) transition and the whole ESR spectrum behaves as an effective spin $S = 1/2$. The Gd$^{3+}$ CEF is then the central issue in the interpretation of the spectra. Nonetheless, in real experiments, a perfect alignment of the magnetic field along the collapsed angle is difficult to achieve, due to field misorientation and/or crystal defects, and always a small misalignment is left in, therefore, $p(\theta) \neq 0$. Thus, even though the observed spectrum presents a single resonance, its linewidth will be always inhomogeneously broadened by CEF effects. This small misalignment would allow to perceive any linewidth anomaly caused by an even tiny CEF effect.

Figure \ref{Fig03}(b) shows the $T$-evolution of the GDCEF used to simulate the $T$-dependence of the nearly collapsed spectra of Figure \ref{Fig03}(a) in the $T$-interval (5.3 K $\leq T \leq$ 12.0 K), which is in Figure \ref{Fig04} denoted by the yellowish band. The experimental $T$-dependence of the ESR spectrum and their respective simulations, with (solid yellow lines) and without (solid red lines) GDCEF, are also presented. Notice the smooth change of the ESR spectra. The linewidth broadens caused by the slow dynamic CEF fluctuation and the lineshape goes from pure Lorentzian to Dysonian one, due to the decrease of the skin depth as a consequence of the conductivity increase in this anomalous linewidth temperature interval \cite{souza2020metallic}. 

Figure \ref{Fig04}(a) shows the $T$-dependence of the X and Q-band Gd$^{3+}$ ESR linewidth near the collapsed angle for {Sm$_{0.9996}$Gd$_{0.0004}$B$_{6}$}. In the absence of $b_{4}$ GDCEF contribution to the linewidth (red solid circles in Fig. \ref{Fig04}(a)), we obtain $E_{ex}$ = 60 K, which is close to the SmB$_6$ hybridization gap \cite{eo2018robustness}. We used $\Delta H_{res}\approx$ 20 Oe, and as expected, there is no Korringa relaxation rate at low-$T$, but a small one, $b = 0.01(2)$ Oe/K, is observed for $T$ above 8 K, consistent with the bulk conductivity \cite{eo2018robustness,souza2020metallic} that sets in near this temperature. 

 In Figure \ref{Fig04}(c) we present the $T$-dependence of the asymmetry parameter, $\xi$, of our ESR results. The ESR resonances of Figure \ref{Fig03}(a) clearly display a change from insulator ($\xi \approx$ 1) to metallic ($\xi \approx$ 0.55) lineshape symmetry, within the temperature interval where anomalous broadening is observed. It was shown that the lineshape evolves from a symmetric Lorentzian  to an asymmetric Dysonian upon temperature increase. This behavior, together with an enhanced conductivity, is associated with a monotonic crossover from insulator-to-metal phases occurring within 6 K $\leq T \leq$ 12 K and provides strong evidence for the local closure of the hybridization gap caused in part by the electron-phonon (rattling) interactions \cite{venegas2016collapse} 

The linewidth behavior according to Eq. (\ref{Eq06}), and showed in Figure \ref{Fig04}(a), describes reasonably well the Gd$^{3+}$ ESR linewidth in the low-$T$ and high-$T$ regimes. However, in the temperature interval of the anomalous linewidth, there is an evident discrepancy between the experimental linewidth and that predicted by the VF model itself. Using VF, Korringa relaxation and a $T$-dependent GDCEF of $b_{4}$, the experimental lineshape and linewidth are both well reproduced in that temperature range, for this particular magnetic field orientation (Figures \ref{Fig03}(a) and \ref{Fig04}(a)). There was no need to consider any Gd$^{3+}$ spin-spin exchange contribution in these simulations, which is further consistent with the extreme low Gd concentration in these single crystals.

It is remarkable that the anomalous linewidth temperature interval coincides with the anomalous impedance regime found in a low-$T$ Impedance Spectroscopy study of SmB$_6$ single crystals that leads to a current controlled negative differential resistance \cite{Jolanta2021}. A possible connection with the parameter $\xi$ is presented in the Appendix A.

As illustrated in Figure \ref{Fig03}(b), $\sigma_{b_4}$ increases dramatically inside the anomalous interval. This behavior is also evident in the inset of this Figure where $\sigma_{b_4}$ rises until stabilizing near $T\approx$ 10 K. For higher-$T$, the contribution of the GDCEF to the inhomogeneous linewidth is overcome by the Korringa and VF narrowing effects. Hence, the ESR linewidth is then consistently described by Eq. (\ref{Eq06}). It is worth mentioning that similar behavior can be also verified in the other field orientations. Notice, however, that the increase of $\sigma_{b_4}$, although evident, is less pronounced when $H$ is applied along [001] and [110]. This is because when $H\parallel 30^\circ$, the spectra is about to collapse and the same Gaussian distribution causes, obviously, at any other orientation, distinct linewidths increase of the individual resonances due to a static inhomogeneous broadening caused by the presence of crystal defects. 

We shall point out that the GDCEF of $b_{4}$ used to simulate the ESR spectra takes into account either positive and negative values of $b_{4}$ (Figures \ref{Fig02}(c),(d) and \ref{Fig03}(b)). Experimentally, it is verified that most insulating materials present negative $b_{4}$ while conductors present positive ones.\cite{duque2007exchange,Barberis1979} To the best of our knowledge, SmB$_{6}$ is the first compound where negative and positive values of $b_{4}$ may coexist at a nanometer scale, with temperature playing the role of the control parameter. For the Kondo insulator SmB$_{6}$, positive and negative values of the $b_{4}$ may indicate the presence of a heterogeneous phase where metallic and insulating regions coexist in this $T$-range  as a consequence of a local loss of coherence between the hybridized $d$-$ce$ and Sm-$4f$ electrons due to the CEF fluctuations. It is worth mentioning that this heterogeneous phase is quite different than that observed for Gd$^{3+}$, Er$^{3+}$ and Eu$^{2+}$ doped CaB$_{6}$ at much higher concentrations, where self doping and bound defects percolate to build up the metallic regions \cite{Urbano2002,Urbano2005}.

In Appendix B we present a dynamic Jahn-Teller-like model explaining the slow time-dependent variations of the cubic CEF parameter $b_4$ giving rise to insulating and metallic values for $b_4$. Superposition of $\Gamma_1$-oscillations modes could yield distributions as in Figs. \ref{Fig03} and \ref{Fig04}. 

Another possible origin for the linewidth anomaly could be a slow $g$-value fluctuation, but this is ruled out by ESR experiments carried out in different microwave frequencies, 9.4 GHz and 34.1 GHz, which, within the accuracy of the experiments, presented the same $\Delta H(T)$ (Figure \ref{Fig04}(a)) \cite{abragam2012electron}.

The $g$-shift, $\Delta g$ = $g_{\rm{exp}}$ - $g_{\rm{ins}}$, calculated with respect to the $g$-value of isolated Gd$^{3+}$ ions in insulators ($g_{ins}$ = 1.993) is presented in Figure \ref{Fig04}(b) \cite{abragam2012electron}. The negative $\Delta g$ measured in X- and Q-bands is nearly $T$-independent and is probably associated with the covalent antiferromagnetic coupling between the hydrogen-like donor bound-state to the Gd$^{3+}$ impurity \cite{duque2007exchange}. Furthermore, Sm$_{1-x}$Gd$_x$B$_6$ samples with $x$ = 0.02 and 0.004 show very similar $\Delta H$ and $\Delta g$ \cite{souza2020metallic}. As a result, the observed linewidth anomaly cannot be attributed to magnetic spin-spin interactions of Gd$^{3+}$ ions. We should also point out that the slow magnetic field dynamics observed by $\mu$SR experiments at low-$T$ and low-$H$ is disregarded here because it vanishes at the temperature and field ranges used in our ESR experiments \cite{biswas2014low,biswas2017suppression}.

Moreover, a weak and broad additional dispersionless mode was observed in the energy gap between the acoustic and optical modes due to the nonadiabatic interaction of phonons \cite{alekseev_2_1989}.
Notice that the small Sm$^{2.6+}$ ions may experience anharmonic rattling vibrations at interstitial positions of the large lattice cage of SmB$_6$ \cite{lesseux_anharmonic_2017}. This is probably the responsible mechanism for the observed dispersionless mode. Thus, since lattice vibrations (phonons) are expected to interfere strongly with the Sm fluctuation valence because of their nearly coincident characteristic times, ($10^{12}$ s$^{-1}$ to $10^{15}$ s$^{-1}$) \cite{alekseev_2_1989,zirngiebl_relation_1986,batkova2017effect}, it is conceivable that slow charge fluctuations, leading to slow CEF fluctuations, result from the interfering beating between these two fast phenomena of close frequencies. Moreover, inelastic neutron scattering experiments demonstrated the existence of magnetic in-gap bound states in SmB$_6$ \cite{alekseev_1_1993,Fuhrman2015}. Besides, two dispersionless magnetic excitation at energies $h\nu_1 \simeq$ 36 and $\simeq$ 14 meV were confirmed within the direct gap of SmB$_6$. Under this scenario, the $T$-dependent high field NMR measurements of the $^{11}$B Knight shift and spin-lattice relaxation rates showed a marked decrease of 1/$T_1$ with increasing $H$ for $T \leq$ 10 K. Hence, the progressive suppression of the spin-lattice relaxation channel upon high fields suggests that the magnetic in-gap bound states do play an important role in this process \cite{caldwell2007high,takigawa1981nmr}. Nevertheless, nuclear relaxation due to slow charge fluctuations of the CEF gradients was disregarded by isotopic effect based on similar $^{10}$B and $^{11}$B NMR results, and the nature of the nuclear spin-lattice relaxation was assumed to be purely magnetic.\cite{pena1981nmr} In contrast to NMR, the ESR results presented in this work did not capture any magnetic effects as revealed by the $T$-independent $g$-shift reported in Figure \ref{Fig04}(a).

\section{\label{sec:conclusion}Conclusion}

In summary, we have performed frequency and $T$-dependent ESR experiments in the Gd$^{3+}$-doped SmB$_{6}$ Kondo insulator compound with $x_{Gd}\simeq$ 0.0004 and magnetic field applied in different crystallographic directions. Based on the interconfigurational VF model, narrowing effects on the Gd$^{3+}$ ESR fine structure at different crystallographic directions was clearly seen (Figures \ref{Fig02}(a) and \ref{Fig02}(b)). For $H$ at nearly the collapsed fine structure, the ESR linewidth follows the expected exponential-like behavior yielding an activation energy of $\simeq$ 60 K, comparable with the Kondo hybridization gap of SmB$_{6}$. However, in the temperature interval of 5.3 K $\leq T \leq$ 12.0 K, an anomalous and inhomogeneous contribution to the ESR linewidth is observed, which was accounted by a $T$-dependent GDCEF of $b_{4}$ parameter in Eq. (\ref{Eq4}). This theoretical framework was able to mimic the slow CEF fluctuations  (slower than $\sim$10 GHz) and reproduce the anomalous linewidth broadening observed in this $T$-interval.

In this $T$-interval a crossover from insulator to metallic environment is observed, and coincides with anomalies of the impedance of SmB$_6$ as is discussed in Appendix A. In Appendix B we present a dynamic Jahn-Teller-like model, by coupling the $^8$S-states of Gd$^{3+}$ to the cage breathing mode, that can explain the slow fluctuations of $b_4$ with positive and negative values of $b_4$.  
This corresponds to fluctuations with time of the crystalline CEF parameter $b_4$. The amplitude of the J-T oscillation increases with temperature. Hence, in Figure \ref{Fig04}(a), $\Delta H$ keeps increasing with $T$ above 12 K. Our simulations indicate coexistence of positive and negative $b_4$ values, suggesting the existence of an heterogeneous phase of metallic and insulating regions. We have shown that our ESR data can be fully explained by assuming a slow CEF fluctuation concomitantly with fast phonon and VF effects. 

Our work introduces a new perspective to the enigmatic properties of SmB$_6$, showing that the interplay between fast VF and phonons give rise to slow charge fluctuations that provide an important clue for the understanding of the physical properties in this SmB$_6$ Kondo system. Nonetheless, further theoretical work is required to fully understand the mechanism involved in the interplay between slow CEF fluctuation, phonons and VF in SmB$_6$. Besides, it is possible that slow CEF fluctuations may be also observed at low-$T$ as a barely perceptible anomaly in the $^{149}$Sm Quadrupole Synchrotron Radiation-Mossbauer spectra \cite{Desmond1996,Tsutsui2016,Tsutsui2019} associated, however, to anharmonic dynamic Jahn-Teller of Sm rattling oscillations \cite{lesseux_anharmonic_2017}, hence, giving further support to the existence of slow CEF fluctuations at low-$T$ in SmB$_6$.

Finally, we believe that our ESR results in the Gd$^{3+}$-doped SmB$_{6}$ Kondo insulator have captured relevant features of this system, $i.e.$, Valence Fluctuation, phonons, charge fluctuations and $T$-dependent hybridization gap \cite{caldwell2007high,takigawa1981nmr} \cite{biswas2014low,biswas2017suppression}  \cite{venegas2016collapse} \cite{alekseev_2_1989,riseborough_9_2000}.

\begin{acknowledgments}
	
	This work was supported and performed under the auspices of FAPESP\ (SP-Brazil) through Grants No 2020/12283-0 2018/11364-7, 2017/10581-1, 2013/17427-7, 2012/04870-7, 2012/05903-6; National Council for Scientific and Technological Development $-$ CNPq Grants No 309483/2018-2, 442230/2014-1, and 304496/2017-0; CAPES and FINEP-Brazil. Work at Los Alamos National Laboratory (LANL) was performed under the auspices of the U.S. Department of Energy, Office of Basic Energy Sciences, Division of Materials Science and Engineering.
	
\end{acknowledgments}

\section{Appendix A}{\bf \underline{Low-$T$ impedance of SmB$_6$}}
\vskip 0.2in

The dielectric response of SmB$_6$ at low-$T$ can be divided into three temperature regimes: $(i)$ for $T < 4$ K only the topological surface states conduct while the bulk states are frozen out, $(ii)$ for $T > 10$ K the conduction is predominantly through the activated semiconductor states, and $(iii)$ in the nontrivial intermediate regime both channels participate in the transport.

The equivalent circuit for the system consists of a resistor $R_b$ and a capacitor $C_b$ connected in parallel to parametrize the insulating bulk and a parallel connection of a resistor $R_s$ and an inductor $L_s$ for the impedance of the surface states. The surface and bulk elements are, of course, connected in parallel. The impedance of SmB$_6$ platelets has recently been measured by Stankiewicz \textit{et al.} \cite{Jolanta2021} in the relevant $T$-range up to 15 K. The voltage and current are out of phase giving rise to Lissajous curves \cite{kim2014topological} in the intermediate region $(iii)$ and for very thin SmB$_6$ film a negative differential resistance leads to self-sustained voltage oscillations at low-$T$ \cite{kim2014topological,Stern2016oscillations,Casas2018heating} as a consequence of Joule heating. The intermediate $T$-region $(iii)$, corresponds approximately to the anomalous linewidth temperature interval in Figure \ref{Fig04}.

The quantity $\xi$ in Fig. \ref{Fig04}(c) parametrizes the crossover from an insulating to a metallic phase and in Fig. \ref{Fig04}(a) the deviation between the experimental X-band measurements and the corresponding CEF simulation in the anomalous linewidth temperature interval. The anomalies in the ESR spectra are therefore related to the anomalies in the impedance of SmB$_6$. The deviations of $\xi$ are consistent with the Joule heating.

\section{Appendix B}{\bf \underline{Coupling to lattice vibrations}}
\vskip 0.2in

In a cubic environment there are vibrations of symmetry $\Gamma_1$, $\Gamma_3$ and $\Gamma_5$ that in principle can couple to an impurity.  The $\Gamma_3$ and $\Gamma_5$ are degenerate and give rise to a Mexican hat, which lifts the degeneracy, while the $\Gamma_1$ mode is non-degenerate. In Ref. \cite{lesseux_anharmonic_2017} we considered the dynamic Jahn-Teller effect of Er$^{3+}$ ions in SmB$_6$ with phonons of $\Gamma_3$ symmetry.  In the case of Gd$^{3+}$ (an S-state) the mode of $\Gamma_1$ symmetry (breathing mode) can couple without breaking the cubic symmetry, but giving rise to a time dependent b$_{4}$. The coupling of the $\Gamma_3$ and $\Gamma_5$ modes give rise to tetragonal and trigonal spin-lattice coefficients, respectively \cite{Calvo1971spinlattice}.

The unperturbed half-filled 4$f^7$ shell of Gd$^{3+}$ has a $^8$S$_{7/2}$ ground state. Perturbations are the spin-orbit interaction and the cubic CEF Hamiltonian.  The perturbation expansion up to fourth order in the spin-orbit coupling ($\xi_{so} = 1480$ cm$^{-1}$) was carried out by Wybourne \cite{Wybourne1966}
\begin{eqnarray}
0.98655 |{^8}_{7}S> &+& 0.16176 |{^6}_{5}P> - 0.01232 |{^6}_{7}D> \nonumber \\
&+& 0.00100 |{^6}_{5}F> - 0.00014 |{^6}_{7}G>  \label{perturb}
\label{Eq7} 
\end{eqnarray}
leading to a reduced $g$-factor of 1.99454. Here the left-hand superscript is $2S+1$ of the multiplet and the left-hand subscript is the seniority number.\cite{Racah1943} The cubic CEF parameter $b_4$ is now obtained as the matrix element of the CEF Hamiltonian with the wave function Eq. (\ref{Eq7}).

The static $b_4$ is given by the point-charge-model for the CEF. In addition one needs to consider the screening of the ionic charges by the $ce$, which reduces the CEF effect.  The breathing mode phonon changes the size of the $B_6$-cage without destroying the symmetry. This gives rise to a modulation of $b_4$ with time and hence to a distribution of $b_4$-values at a given time.  If this modulation is large enough the system could have some regions with negative $b_4$ as for an insulator and others with positive $b_4$ as for a conductor. Since the breathing mode is larger than the size of a unit cell, it is to be expected that it arises from the acoustic branches in SmB$_6$.  Hence, the time-dependence of the modulation of b$_4$ is rather slow.

\bibliography{Paper-SmB6V2}

\end{document}